\documentstyle[12pt]{article}
\def\la{\mathrel{\mathchoice {\vcenter{\offinterlineskip\halign{\hfil
$\displaystyle##$\hfil\cr<\cr\sim\cr}}}
{\vcenter{\offinterlineskip\halign{\hfil$\textstyle##$\hfil\cr<\cr\sim\cr}}}
{\vcenter{\offinterlineskip\halign{\hfil$\scriptstyle##$\hfil\cr<\cr\sim\cr}}}
{\vcenter{\offinterlineskip\halign{\hfil$\scriptscriptstyle##$\hfil\cr<\cr\sim
\cr}}}}}

\begin{document}
\begin{center}
{\bf Probing deviations from tri-bimaximal mixing through ultra high 
energy neutrino signals} 
\end{center}
\begin{center}
Debasish Majumdar and Ambar Ghosal
\end{center}
\begin{center}
{\it Saha Institute of Nuclear Physics} \\
{\it 1/AF Bidhannagar, Kolkata 700 064, INDIA}
\end{center} 
\begin{center}
{\bf Abstract}
\end{center}
We investigate deviation from the tri-bimaximal mixing in the case of 
ultra high energy neutrino using ICECUBE detector. We consider the ratio of 
number of muon tracks to the shower generated due to eletrons and hadrons.
Our analysis shows that for tri-bimaximal mixing the ratio comes out 
around 4.05. Keeping $\theta_{12}$ and $\theta_{23}$ fixed at tri-bimaximal 
value, we have varied the angle $\theta_{13}$ = $3^o$, $6^o$, $9^o$ 
and the value of the ratio gradually decreases. The variation of ratio 
lies within $8\%$ to $18\%$ from the tri-bimaximal mixing value and it is 
very difficult to detect such small variation by the ICECUBE detector. 
\newpage
\section{Introduction}

Various experiments for solar and atmospheric neutrinos provide 
a range for the values of solar mixing angle $\theta_\odot = \theta_{12}$ 
(the 1-2 mixing angle) \cite{sfit}
that corresponds to solar neutrino oscillations and also a range for 
atmospheric mixing 
angle $\theta_{\rm atm} = \theta_{23}$ (the 2-3 mixing angle) \cite{sfit} 
around their best fit values. The tri-bimaximal mixing condition 
of neutrinos are given by $\sin \theta_{12} = \frac {1} {\sqrt {3}}$,
$\sin \theta_{23} = \frac {1} {\sqrt {2}}$ and $\sin \theta_{13} = 0$
\cite{tbi}. Possible deviations from tri-bimaximal mixing can be 
obtained by probing the ranges of $\theta_{12}$ and $\theta_{23}$
given by the experiments. Also the exact 13 mixing angle 
$\theta_{13}$ is not known except that the CHOOZ \cite{chooz} gives 
an upper limit for $\theta_{13} (< 9^o)$. Probing the deviations 
of $\theta_{12}$ and $\theta_{23}$ for different values 
for $\theta_{13}$ is significant not only to understand the 
neutrino flavour oscillations in general but also for the purpose 
of model building for neutrino mass matrices. 

In this work we explore the possibility for ultra high energy (UHE) 
neutrinos from distant Gamma Ray Bursts (GRBs) 
for probing the signatures of these deviations of the values of the mixing 
angles from tri-bimaximal mixing as discussed above. One such proposition
of using UHE neutrinos is described in a recent work by Xing \cite{xing}. 
Gamma Ray Bursts are short lived but intense burst of gamma rays.
During its occurence it outshines all other luminous objects in the 
sky. Although the exact machanism of GRBs could not be ascertained 
so far but the general wisdom is that it is powered by a central 
engine provided by a failed star or supernova that possibly turned into a 
black hole, accretes mass at its surroundings. This infalling mass 
due to gravity bounces back from the surface of black hole much the same way 
as the supernova explosion mechanism and a shock is generated that 
flows radially 
outwards with enormous amount of energies ($\sim 10^{53}$ ergs). This 
highly energetic shock wave drives the mass outwards, in the form 
of a ``fireball" that carries in it, protons, $\gamma$ etc. The pions 
are produced when the accelerated protons inside the fireball 
interacts with $\gamma$ through a cosmic beam dump 
process. UHE neutrinos are produced by the decay of these pions.
Thus a generic cosmic accelerator accelerates the protons 
into very high energies which then beam dump on $\gamma$ in the ``fireball"
as also at the cosmic microwave background (CMB) and ultra high energy 
neutrinos are produced.     

The GRB neutrinos, due to their origin at astronomical distances from earth, 
provide a very long baseline for the earth bound detectors for UHE 
neutrinos such as ICECUBE \cite{icecube}. The oscillatory part of the neutrino 
flavour oscillation probabilities ($\sin^2(\Delta m^2 [L/4E])$) averages 
out to 1/2 becuase of this very long baseline $L$ ($\sim$ hundreds of Mpc)
and the $\Delta m^2$ (mass square difference of two neutrinos) 
range obtained from solar and atmospheric neutrino
experiments are $\Delta m^2_{21} \sim 10^{-4}$ eV$^2$ and 
$\Delta m^2_{32} \sim 10^{-3}$ eV$^2$ respectively ($L/\Delta m^2 >> 1$). 
Thus for neutrino flavour
oscillation, in this case, the effect of $\Delta m^2$ is washed out and 
governed only by the three mixing angles namely $\theta_{12} = \theta_\odot$,
$\theta_{23} = \theta_{\rm atm}$ and $\theta_{13}$. The purpose of the 
present work is to probe whether or not the possible variations 
of $\theta_{12}$ and $\theta_{23}$  from their best fit values can be 
ascertained by UHE from distant GRBs.  

The GRB neutrinos, on arriving the earth, undergo charged current (CC) and 
neutral current (NC) interactions with the earth rock and the detector 
material. The CC interactions of $\nu_\mu$ produce secondary 
muons and the same  
for electrons produce electromagnetic shower ($\nu_\mu + N \rightarrow 
\mu + X$ and $\nu_e + N \rightarrow e + X$). The former will produce   
secondary muon tracks and can be detected by track-signal produced  
by the Cerenkov light emitted by these muons during their passage 
through a large underground water/ice Cerenkov detectors like ICECUBE.
The ICECUBE is a 1km$^3$ detector in south pole ice and can be considered
to be immersed in the target material for the UHE neutrinos where 
the neutrino interactions are initiated.   
In case of $\nu_e$, the electrons from the $\nu_e N$ CC interactions, 
shower quickly and can also be detected by such ICECUBE detector.   
The case of $\nu_\tau$ is somewhat complicated. The first
CC interaction of $\nu_\tau$ ($\nu_\tau + N \rightarrow \tau + X$) 
produces a shower (``first bang") alongwith a $\tau$ track. But the 
$\nu_\tau$ is regenerated (with diminished energy) by the decay of $\tau$ 
and in the process produces another hadronic or electromagnetic 
shower (``second bang"). The whole process is called  
double bang event. In case the first
bang could not be detected, then by possible detection of 
second bang (with showers) the $\tau$ track can be reconstructed or 
identified and 
this scenario (the $\tau$ track and the second bang) is 
called the lollipop events. An inverted lollipop event is one where 
only the first bang ($\nu_\tau + N \rightarrow \tau + X$)  is detected 
and the subsequent $\tau$ track is detected or reconstructed. As mentioned 
in Ref. \cite {cao}, the detection of $\nu_\tau$ from their CC interaction 
mentioned above is not every efficient by a 1km$^3$ detector since 
the double bang events can possibly be detected only for the $\nu_\tau$ 
energies between 1 PeV to 20 PeV beyond which the tau decay length 
is longer than the width of such detector and at still higher energies 
the flux is too small for such detectors for their detection. 
Hence, in the present work we do not consider the events initiated 
by $\nu_\tau N$ CC interactions. However, for $\nu_\tau$ we consider 
the process that may yield events higher than the ``double bang" events.
We consider the decay channel of $\tau$ lepton \cite{sarada},  
obtained from charged current interactions of $\nu_\tau$, where muons 
are produced 
($\nu_\tau \rightarrow \tau \rightarrow {\bar \nu_\mu} \mu \nu_\tau$)
which can then be detected as muon tracks \cite{nayantau}
in ICECUBE detector. 
The neutral current (NC) interactions of all flavours however will 
produce the shower events at ICECUBE and they are considered in this 
investigation.

This paper is organised as follows. In Section 2 we describe the formalism 
for neutrino fluxes of the three species while reaching the earth. The 
nature of the GRB flux taken for present calculations is also discussed. 
The flux suffers flavour oscillations while traversing from GRB site 
to the earth. The oscillation probabilities are also calculated and 
the oscillated flux obtained on reaching the earth is determined.
They are given in Section 2.1.  
We also describe in this section the analytical expressions for the yield of 
secondary muons and shower events at the ice Cerenkov kilometre square
detector like ICECUBE. This is given in Section 2.2. The actual calculations
and results are discussed in Section 3. Finally, in Section 4,  
some discussions and summary are given.

\section{Formalism}

\subsection{GRB Neutrinos Fluxes}

The neutrino production in GRB is initiated through the process 
of cosmological beam dump by which a highly accelerated protons
from GRB interacts with $\gamma$ to produce pions which in turn 
decays to produce $\nu_\mu ({\bar {\nu_\mu}})$ and $\nu_e ({\bar {\nu_e}})$
much the same ways as atmospheric neutrinos are produced. They are 
produced in the proportion $2\nu_\mu : 2{\bar {\nu_\mu}} : 1\nu_e :
1{\bar {\nu_e}}$ \cite{raj1}. 

For the present calculation we consider the isotropic flux 
\cite{waxman} resulting from the summation over the sources and 
as given in Gandhi et al \cite{raj2}.  The isotropic GRB flux for 
$\nu_\mu + {\bar {\nu_\mu}}$ is given as 
\begin{equation}
{\cal F}(E_\nu) = \frac {d N_{\nu_\mu + {\bar {\nu_\mu}}}} {dE_\nu} = 
{\cal N} \left ( \frac {E_\nu} {1{\rm GeV}} \right )^{-n} 
{\rm cm}^{-2} {\rm s}^{-1} {\rm sr}^{-1} {\rm GeV}^{-1}
\end{equation}
In the above,
$$
{\cal N} = 4.0 \times 10^{-13},\,\,\,\, n=1,\,\,\,\, {\rm for}\,\,\, E_\nu < 10^5
\,\,{\rm GeV} 
$$
$$
{\cal N} = 4.0 \times 10^{-8},\,\,\,\, n=2,\,\,\,\, {\rm for} \,\,\,E_\nu > 10^5
\,\,{\rm GeV} 
$$
Thus, 
\begin{equation}
\begin{array}{rclclclcl}
\displaystyle \frac {d N_{\nu_\mu}} {dE_\nu} &=& \phi_{\nu_\mu} &=&
\displaystyle \frac {d N_{\bar {\nu_\mu}}} {dE_\nu} &=& \phi_{\bar {\nu_\mu}}
&=& 0.5{\cal F}(E_\nu) \\
&&&&&&&& \\
\displaystyle \frac {d N_{\nu_e}} {dE_\nu} &=& \phi_{\nu_e} &=& 
\displaystyle \frac {d N_{\bar {\nu_e}}} {dE_\nu} &=& \phi_{\bar {\nu_e}} &=&
0.25{\cal F}(E_\nu)
\end{array} 
\end{equation}
The neutrinos undergo flavour oscillation during their passage from the 
GRB to the earth. Under three flavour oscillation, the $\nu_e$ and $\nu_\mu$
originally created at GRB will be oscillated to $\nu_\tau$.  
Thus after flavour oscillations, the $\nu_e$ fluxes ($F_{\nu_e}$),
$\nu_\mu$ fluxes ($F_{\nu_\mu}$), $\nu_\tau$ fluxes ($F_{\nu_\tau}$)  
become 
\begin{eqnarray}
F_{\nu_e} &=& P_{\nu_e \rightarrow \nu_e} \phi_{\nu_e} + 
P_{\nu_\mu \rightarrow \nu_e} \phi_{\nu_\mu} \nonumber \\
F_{\nu_\mu} &=& P_{\nu_\mu \rightarrow \nu_\mu} \phi_{\nu_\mu} +
P_{\nu_e \rightarrow \nu_\mu} \phi_{\nu_e} \nonumber \\
F_{\nu_\tau} &=& P_{\nu_e \rightarrow \nu_\tau} \phi_{\nu_e} +
P_{\nu_\mu \rightarrow \nu_\tau} \phi_{\nu_\mu}\,\,\, . 
\end{eqnarray}

The transition probability of a neutrino of flavour $\alpha$ to a flavour
$\beta$ is given by,
\begin{equation}
P_{\nu_\alpha \rightarrow \nu_\beta}
= \delta_{\alpha\beta} - 4\displaystyle\sum_{j>i} U_{\alpha_i}U_{\beta_i}
U_{\alpha_j}U_{\beta_j} \sin^2 \left ( \frac {\pi L} {\lambda_{ij}} \right )
\end{equation}
In the above oscillation length
$\lambda_{ij}$ is given by
\begin{equation}
\lambda_{ij} = 2.47\,\, {\rm Km} \left ( \frac {E} {\rm GeV} \right )
\left ( \frac {{\rm eV}^2} {\Delta m^2} \right )
\end{equation}
Because of astronomical baseline $\Delta m^2L/E >> 1$,
the oscillatory part
becomes averaged to half. Thus,
\begin{equation}
\left \langle \sin^2 \left ( \frac {\pi L} {\lambda_{ij}} \right )
\right \rangle = \frac {1} {2}
\end{equation}
Therefore
\begin{eqnarray}
P_{\nu_\alpha \rightarrow \nu_\beta}
&=& \delta_{\alpha\beta} - 2\displaystyle\sum_{j>i} U_{\alpha_i}U_{\beta_i}
U_{\alpha_j}U_{\beta_j} \nonumber \\
&=& \delta_{\alpha\beta} - \displaystyle\sum_i U_{\alpha_i}U_{\beta_i}
\left [ \displaystyle\sum_{j \neq i} U_{\alpha_j}U_{\beta_j} \right ]
\nonumber \\
&=& \displaystyle\sum_j |U_{\alpha_j}|^2 |U_{\beta_j}|^2
\end{eqnarray}
where use has been made of the condition $\sum_i U_{\alpha_i}U_{\beta_i} =
\delta_{\alpha\beta}$.

With Eq. (7), Eq. (3) can be rewritten in matrix form

\begin{eqnarray}
\left ( \begin{array}{c} F_{\nu_e} \\ F_{\nu_\mu} \\ F_{\nu_\tau} \end{array}
\right ) 
&=& 
\left ( \begin{array}{ccc} U_{e1}^2 & U_{e2}^2 & U_{e3}^2 \\
U_{\mu 1}^2 & U_{\mu 2}^2 & U_{\mu 3}^2 \\
U_{\tau 1}^2 & U_{\tau 2}^2 & U_{\tau 3}^2 \end{array} \right ) 
\left ( \begin{array}{ccc} U_{e1}^2 & U_{\mu 1}^2 & U_{\tau 1}^2 \\
U_{e2}^2 & U_{\mu 2}^2 & U_{\tau 2}^2 \\
U_{e3}^2 & U_{\mu 2}^2 & U_{\tau 3}^2 \end{array} \right )  
\left ( \begin{array}{c} \phi_{\nu_e} \\ \phi_{\nu_\mu} \\ \phi_{\nu_\tau} 
\end{array} \right ) \nonumber \\
&=& \left ( \begin{array}{ccc} U_{e1}^2 & U_{e2}^2 & U_{e3}^2 \\
U_{\mu 1}^2 & U_{\mu 2}^2 & U_{\mu 3}^2 \\
U_{\tau 1}^2 & U_{\tau 2}^2 & U_{\tau 3}^2 \end{array} \right )
\left ( \begin{array}{ccc} U_{e1}^2 & U_{\mu 1}^2 & U_{\tau 1}^2 \\
U_{e2}^2 & U_{\mu 2}^2 & U_{\tau 2}^2 \\
U_{e3}^2 & U_{\mu 2}^2 & U_{\tau 3}^2 \end{array} \right )
\left ( \begin{array}{c} 1 \\ 2 \\ 0
\end{array} \right ) \phi_{\nu_e}
\end{eqnarray}
In Eq. (8) above, we have used the initial flux ratio from GRB to be 
$\phi_{\nu_e} : \phi_{\nu_\mu} : \phi_{\nu_\tau} = 1 : 2 : 0$. From Eq. (8)
it then follows that,
\begin{eqnarray}
F_{\nu_e} & = & \left \{ U_{e1}^2 [1 + (U_{\mu 1}^2 - U_{\tau 1}^2)] + 
          U_{e2}^2 [1 + (U_{\mu 2}^2 - U_{\tau 2}^2)] + \right . \nonumber \\
  && \left .  U_{e3}^2 [1 + (U_{\mu 3}^2 - U_{\tau 3}^2)] \right \} 
\phi_{\nu_e}  \nonumber \\
&& \nonumber \\
F_{\nu_\mu} & = & \left \{  U_{\mu 1}^2 [1 + (U_{\mu 1}^2 - U_{\tau 1}^2)] + 
       U_{\mu 2}^2 [1 + (U_{\mu 2}^2 - U_{\tau 2}^2)] + \right . \nonumber \\
 && \left .  U_{\mu 3}^2 [1 + (U_{\mu 3}^2 - U_{\tau 3}^2)] \right \} 
\phi_{\nu_e}   \nonumber \\
&& \nonumber \\
F_{\nu_\tau} & = & \left \{ U_{\tau 1}^2 [1 + (U_{\mu 1}^2 - U_{\tau 1}^2)] + 
  U_{\tau 2}^2 [1 + (U_{\mu 2}^2 - U_{\tau 2}^2)] + \right . \nonumber \\
 && \left .  U_{\tau 3}^2 [1 + (U_{\mu 3}^2 - U_{\tau 3}^2)] \right \} 
\phi_{\nu_e}
\end{eqnarray}
The MNS mixing matrix $U$ for 3-flavour case is given as 
\begin{equation}
U = \left ( \begin{array}{ccc} 
c_{12}c_{13} & s_{12}c_{13} & s_{13} \\
-c_{23}s_{12} - s_{23}s_{13}c_{12} & c_{23}c_{12} - s_{23}s_{13}s_{12} &
s_{23}c_{13} \\
s_{23}s_{12} - c_{23}s_{13}c_{12} & -s_{23}c_{12} - c_{23}s_{13}s_{12} &
c_{23}c_{13} 
\end{array} \right )
\end{equation}
We are not considering any CP violation here. Hence Eqs. (3) - (9) above 
also hold for antineutrinos. 

\subsection{Detection of GRB neutrinos}

The $\nu_\mu$'s from a GRB can be detected from the tracks of 
the secondary muons produced through the $\nu_\mu$ CC interactions. 

The total number of secondary muons induced by GRB neutrinos at a
detector of unit area is given by (following \cite{gaisser,raj1,nayan})
\begin{equation}
S = \int_{E_{\rm thr}}^{E_{\nu{\rm max}}} 
dE_\nu \frac {dN_{\nu}} {dE_\nu} P_{\rm surv}(E_\nu)
P_\mu(E_\nu,E_{\rm thr})
\end{equation}
In the above, $P_{\rm surv}$ is the probability that a neutrino reaches 
the detector without being absorbed by the earth. This is a function of 
the neutrino-nucleon interaction length in the earth and the effective 
path length $X(\theta_z)$ (gm cm$^{-2}$) for incident neutrino 
zenith angle $\theta_z$
($\theta_z = 0$ for vertically downward entry with respect to the detector).
This attenuation of neutrinos due to passage through the earth is 
referred to as shadow factor. For an isotropic distribution of flux,
this shadow factor (for upward going neutrinos) is given by
\begin{equation}
P_{\rm surv}(E_\nu) = \frac {1} {2\pi} 
\int^0_{-1} d\,\,\cos \theta \int d\phi {\rm exp}[-X(\theta_z)/L_{\rm int}].
\end{equation}
where interaction length $L_{\rm int}$ is given by 
\begin{equation}
L_{\rm int} =  \frac {1} {\sigma^{\rm tot}(E_\nu) N_A}
\end{equation}
In the above $N_A (= 6.022 \times 10^{23} {\rm gm}^{-1})$ is the Avogadro number
and $\sigma^{\rm tot} (= \sigma^{\rm CC} + \sigma^{\rm NC})$ is the 
total cross section. The effective path length $X(\theta_z)$ is 
calculated as 
\begin{equation}
X(\theta_z) = \int \rho (r(\theta_z, \ell)) d\ell.
\end{equation}
In Eq. (9), $\rho (r(\theta_z, \ell)$ is the matter density inside the earth
at a distance
$r$ from the centre of the earth for neutrino path length $\ell$ entering
into the earth with a zenith angle $\theta_z$.
The quantity $P_\mu(E_\nu,E_{\rm thr})$ in Eq. (6) is the 
probability that a secondary
muon is produced by CC interaction of $\nu_\mu$ and reach the 
detector above the threshold energy $E_{\rm thr}$. This is then a 
function of $\nu_\mu N$ (N represents nucleon) - CC interaction 
cross section $\sigma^{\rm CC}$
and the range of the muon inside the rock. 
\begin{equation}
P_\mu(E_\nu,E_{\rm thr}) = N_A \sigma^{\rm CC} 
\langle R(E_\nu;E_{\rm thr})
\rangle
\end{equation}
In the above $\langle R(E_\nu;E_{\rm thr})\rangle$ 
is the average muon range given by
\begin{equation}
\langle R(E_\nu;E_{\rm thr}) \rangle = \frac {1} {\sigma^{\rm CC}}
\displaystyle\int_0^{1 - E_{\rm thr}/E_\nu} 
dy R(E_\nu (1 - y), E_{\rm thr})
\frac {d\sigma^{\rm CC}(E_\nu,y)} {dy}
\end{equation}
where $y = (E_\nu - E_\mu)/E_\nu$ is the 
fraction of energy loss 
by a neutrino of energy $E_\nu$ in the charged current production of
a secondary muon of energy $E_\mu$. Needless to say that a muon thus produced
from a neutrino with energy $E_\nu$ can have the 
detectable energy range between
$E_{\rm thr}$ and $E_\nu$. The range $R (E_\mu, E_{\rm thr})$ for a muon
of energy $E_\mu$ is given as
\begin{equation}
R (E_\mu, E_{\rm thr}) = \displaystyle\int^{E_\mu}_{E_{\rm thr}} 
\frac {dE_\mu} {\langle dE_\mu/dX \rangle} \simeq \frac {1} {\beta}
\ln \left ( \frac {\alpha + \beta E_\mu} {\alpha + \beta E_{\rm thr}}.
\right )
\end{equation}
The average lepton energy loss with energy $E_\mu$ per unit distance 
travelled is given by 
\cite{gaisser} 
\begin{equation}
\left \langle \frac {dE_\mu} {dX} \right\rangle = -\alpha - \beta E_\mu
\end{equation}
The values of $\alpha$ and $\beta$ used in the present calculations
are 
\begin{eqnarray}
\alpha &=& \{ 2.033 + 0.077\ln[E_\mu {\rm (GeV)}] \}\times 10^{-3} {\rm GeV}
{\rm cm}^2 {\rm gm}^{-1} \nonumber \\
\beta &=& \{ 2.033 + 0.077\ln[E_\mu {\rm (GeV)}] \} \times 10^{-6}
{\rm cm}^2 {\rm gm}^{-1} 
\end{eqnarray}
for $E_\mu \la 10^6$ GeV \cite{dar} and 
\begin{eqnarray}
\alpha &=& 2.033 \times 10^{-3} {\rm GeV} 
{\rm cm}^2 {\rm gm}^{-1} \nonumber \\
\beta &=& 3.9 \times 10^{-6} 
{\rm cm}^2 {\rm gm}^{-1}
\end{eqnarray}
otherwise \cite{guetta1}.  
For muon events  obtained from $\nu_\mu$ CC interactions, 
$\frac {d N_\nu}{d E_\nu}$ in Eq. (11)
will be replaced by $F_{\nu_\mu}$ (Eq. 9).

As discussed earlier, the events due to $\nu_\tau$ CC interactions 
is considered only for the process where the decay of secondary 
$\tau$ lepton produces muon which then detected by the muon track. 
The probability of production of muons in the decay channel 
$\tau \rightarrow {\bar \nu_\mu} \mu \nu_\tau$ is 0.18 \cite{sarada,nayantau}.
The generated muon carries a fraction 0.3 of energy of original $\nu_\tau$
(a fraction 0.75 of the energy of the $\nu_\tau$ is carried by secondary 
$\tau$ lepton and a fraction of 0.4 of $\tau$ lepton energy is carried 
by the muon \cite{sarada,gaisser,nayantau}). For the detection of such muons,
the Eqs. (10 - 16) is applicable with properly incorporating the 
muon energy described above. Needless to say, in this case, 
$\frac {d N_\nu}{d E_\nu}$ in Eq. (11)
is to be replaced by $F_{\nu_\tau}$ (Eq. 9).   

For the case of showers, we do not have the advantage of a specific 
track and then the whole detector volume is to be considered. 
The event rate for the shower case is given by 
\begin{equation}
N_{\rm sh} = \int dE_\nu \frac {dN_{\nu}} {dE_\nu} P_{\rm surv}(E_\nu)
\times \int \frac {1} {\sigma^j} \frac {d {\sigma^j}} {dy} P_{\rm int}
(E_\nu,y)\,\,.
\end{equation} 
In the above, $\sigma^j = \sigma^{\rm CC}$ (for electromagnetic shower 
from $\nu_e$ charged current interactions) or $\sigma^{\rm NC}$ as the 
case may be. In the above $P_{\rm int}$ is the probability that a 
shower produced by the neutrino interactions will be detected and is given by 
\begin{equation}
P_{\rm int} = \rho N_A \sigma^j L
\end{equation}
where $\rho$ is the density of the detector material and $L$ is the 
length of the detector (L = 1 Km for ICECUBE).  

For each case of shower events, $\frac {dN_{\nu}} {dE_\nu}$ in Eq. (21)
is to be replaced by $F_{\nu_e}$ or $F_{\nu_\mu}$ or $F_{\nu_\tau}$ as the case 
may be. 

\section{Calculations and Results}

The secondary muon yield at a kilometre scale detector such as 
ICECUBE is calculated using Eqs. (6 - 20). The earth matter density 
in Eq. (9) is taken from \cite{raj1} following the Preliminary 
Earth Reference Model (PREM). The interaction cross-sections - both charged 
current and total - used in these equations are taken from the 
tabulated values (and the analytical form) given in Ref. \cite{raj2}.
In the present calculations $E_{\nu{\rm max}} = 10^{11}$ GeV and 
threshold energy $E_{\rm thr} = 1$ TeV are considered.  

For our investigations, we first define a ratio ${\cal R}$ of the muon   
events (both from $\nu_\mu$ (and ${\bar \nu_\mu}$) and $\nu_\tau$ 
(and ${\bar \nu_\tau}$)) and the shower events. As described in the 
previous sections, the muon events are from $\nu_\mu$ 
(and ${\bar \nu_\mu}$) and $\nu_\tau$ (and ${\bar \nu_\tau}$), whereas
the shower events include electromagnetic shower initiated by CC
interaction of $\nu_e$ and NC interactions of neutrinos of all flavours. 
Therefore,
\begin{equation}
{\cal R} = \displaystyle \frac {T_\mu} {T_{\rm sh}}
\end{equation}
where,
\begin{eqnarray}
T_\mu &=& S({\rm for }\,\,\, \nu_\mu) +  S({\rm for }\,\,\, \nu_\tau) 
\nonumber \\
T_{\rm sh} &=& N_{\rm sh} ({\rm for }\,\,\, \nu_e {\rm\,\, CC\,\,\, interaction}) 
\nonumber \\
        &+& N_{\rm sh} ({\rm for }\,\,\, \nu_e {\rm\,\, NC\,\,\, interaction}) \nonumber \\
        &+& N_{\rm sh} ({\rm for }\,\,\, \nu_\mu {\rm\,\, NC\,\,\, interaction}) \nonumber \\
        &+& N_{\rm sh} ({\rm for }\,\,\, \nu_\tau {\rm\,\, NC\,\,\, interaction})
\end{eqnarray}

The purpose of this work is to explore whether UHE neutrinos from GRB 
will be able to distinguish any variation of $\theta_{12}$ and 
$\theta_{23}$ from their best fit values. The tri-bimaximal mixing 
condition is denoted by the best fit values of $\theta_{12}$ and 
$\theta_{23}$ for $\theta_{13} = 0^o$. The best fit value of  
$\theta_{12} = 35.2^o$ and  that of $\theta_{23} = 45^o$. 
We first vary $\theta_{12}$ in the limit $30^o \leq \theta_{12} \leq 38^o$
and vary $\theta_{23}$ in the limit $38^o \leq \theta_{12} \leq 54^o$
with $\theta_{13} = 0$ and for each case calculate the ratio ${\cal R}$ 
using Eqs. (1 - 24). We find that ${\cal R}$ varies from 
3.14 to 4.25. One readily sees that the variation in muon to shower 
ratio is not very significant. The flux and other uncertainties of the 
detector may wash away this small variations. ${\cal R}$ obtained from 
tri-bimaximal condition given above is 4.05. 

The same operation is repeated for three different values of $\theta_{13}$,
namely $\theta_{13} = 3^o\,\,6^o$ and $9^o$ with similar results. 
The results are tabulated below.

\begin{center}
\begin{tabular}{|c|c|c|c|}
\hline
$\theta_{13}$ & ${\cal R}_{\rm Max}$ & ${\cal R}_{\rm Min}$ & ${\cal R}$ at \\
&&& $\theta_{12} = 35.2^o$, $\theta_{23} = 45^o$ \\
\hline
$0^o$ & 4.78 & 3.80 & 4.05 \\
\hline
$3^o$ & 4.75 & 3.77 & 4.01 \\
\hline
$6^o$ & 4.72 & 3.75 & 3.98 \\
\hline
$9^o$ & 4.69 & 3.73 & 3.96 \\
\hline
\end{tabular}
\end{center}
\noindent {\small Table 1. Maximum and minimum values of ratio ${\cal R}$
for different values of mixing angles} 

We have also plotted the variation of ${\cal R}$ with $\theta_{12}$ and 
$\theta_{23}$ for four fixed values of $\theta_{13}$ as given in Table 1.
These are shown in Figs 1a - 1d for $\theta_{13} =$
$0^o,\,\,3^o\,,\,6^o$ and $9^o$ respectively.

As is evident from Table 1 and Fig. 1, the variation of muon tracks to 
shower ratio is not very significant 
with the deviation from the best fit values of the mixing angles. 
The ratio ${\cal R}$ varies upto only $\sim 18\%$. We have also 
calculated the muon track signal for 1 year of ICECUBE run. 
For $\theta_{13} = 0$, this varies from $\sim 99$ to $\sim 115$, 
whereas the muon yield obtained for tri-bimaximal mixing is 103. 
So the variation for deviation from tri-bimaximal mixing 
condition is between $4\%  - 11\%$. This variation is also not 
significant given the sources of uncertainty in the flux and 
the sensitivity of the ICECUBE detector. 
Firstly, the flux itself can be uncertain 
by several factors. This can induce errors in calculation of muon 
yield and shower rate. If the flux uncertainties are energy-dependent,
even the ratio ${\cal R}$ can also be affected. Also the simulation results 
for ICECUBE detector by Ahrens et al \cite{icecubesim1} shows 
the cosmic neutrino signal is well below the atmospheric neutrino 
background for one year data sample after applying suitable cuts 
(for the source flux $E_\nu^2 \times  dN_\nu/dE_\nu$  
$=10^{-7} {\rm cm}^2 {\rm s}^{-1} {\rm sr}^{-1} {\rm GeV}$). 
The diffuse flux needed 
for a $5\sigma$ significance detection after 1 year is well below 
the experimental limits \cite{icecubesim1,icecubesim2}. There can  
also be systematic uncertainty arises out of optical module (OM) sensitivity
which is affected by the refrozen ice around OM, optical properties of the 
surrounding ice, trapped air bubbles in the OM neighbourhood etc. 
An estimation of these uncertainties for a $E^{-2}$ signal is calculated 
to be around 20\% \cite{icecubesim1}. Taking into account these uncertainties
and sensitivity limit, it is      
difficult by a detector like ICECUBE to detect the deviation ($\la 18\%$),
if any, from tri-bimaximal mixing through the detection of UHE neutrinos from 
a GRB.

\section{Summary and Discussions}

In summary, we investigate the deviation from the well known tri-bimaximal
mixing in the case of Ultra High Energy neutrinos from a Gamma Ray Burst
detected in a kilometer scale detector such as ICECUBE. We have calculated
the ratio ${\cal R}$ of the muon track events and shower events 
(electromagnetic shower from charged current interactions of 
$\nu_e$ and hadronic showers from neutral current interactions of 
neutrinos of all flavours) for tri-bimaximal mixing condition given by 
$\theta_{12} = 35.2^o$, $\theta_{23} = 45.0^o$, $\theta_{13} = 0^o$.
We then investigate the possible variation of  ${\cal R}$ from 
tri-bimaximal mixing condition by varying $\theta_{12}$ and $\theta_{23}$
within their experimentally obtained range for four different values 
of $\theta_{13}$ namely $0^o,\,\,3^o\,,\,6^o$ and $9^o$. 

The isotropic flux of GRB neutrinos are obtained following Waxman-Bahcall 
\cite{waxman} type
parametrization of the flux and summation over the sources. The initial 
parametrization of neutrino flux can be written as  
\begin{equation}
\displaystyle\frac {dN_\nu} {dE_\nu} = \left \{
\begin{array}{c} \frac {A} {E_\nu E_\nu^b}\,\, , \,\,\, E_\nu < E_\nu^b \\
                 \frac {A} {E_\nu^2}\,\, , E_\nu > E_\nu^b
\end{array} \right .
\end{equation}
where $E_\nu^b$ is the spectral break energy ($\sim 10^5$ GeV) and is related 
to photon spectral break energy, Lorentz factor etc.

The GRB neutrinos after reaching the earth has to pass through the 
earth rock (for upward going events) to reach the detector to produce 
muon tracks or shower. In the calculation therefore, 
the attenuation of neutrinos through the earth (shadow factor) is estimated.
The muons produced out of charged current interactions of neutrinos 
should also survive to enter the detector and produce tracks. Therefore,
to estimate the muon track events, the energy loss of muons through the 
rock is also estimated. The average lepton energy loss 
rate (with lepton energy $E_\mu$) 
due to ionisation and the losses due to Bremsstrahlung, pair-production,
hadron production etc. (catastrophic losses) is parametrized as 
$$
\displaystyle \left \langle \frac {dE_\mu} {dX} \right \rangle 
= -\alpha - \beta E_\mu
$$
where $\beta$ describes the catastrophic loss which dominate over the 
ionisation loss above a certain critical energy $\zeta = \alpha/\beta$.
This induces a logarithmic dependence of the lepton energy loss.     

The calculated ratio
${\cal R}$ varies between $\sim 8\%$ to $\sim 18\%$ for variation from 
tri-bimaximal mixing scenario and for different values of $\theta_{13}$. 
Given the sensitivity of the ICECUBE detector in terms of detecting 
GRB neutrino flux and considering other uncertainties 
like that in estimating the flux itself, the atmospheric background, 
low signal yield and the systematic uncertainties  
of the detector, it appears that ICECUBE with its present 
sensitivity will not be able to detect significantly such a small variation 
due to deviations from tri-bimaximal mixing. 
Hence to
detect such small deviation, very precise mesurement is
called for. This requires more data (more years of run)
and larger detector size for more statistics. The increase
in detector size will not widen the deviation of the ratio
significantly as the total area factor of the detector
cancels out in the ratio (Eq. 23) although the total number of both muon
tracks and total shower yield increase significantly.
For the case of shower, the whole detector
volume is to be considered and from Eq. (22), there is
indeed an $L$ dependence. This makes the deviation of
the ratio wider although very marginally as we increase
the detector dimension.
                                                                                
It is difficult to predict the detector dimension and/or
the time of exposure that will be suitable for
such a precision measurement discussed above.
Detailed simulation studies
taking into account factors like atmospheric neutrino background,
photomultiplier tube efficiency and other possible uncertainties
like the one carried out in Ref. [16] is required for being able to comment
on the detector parameters for such precise measurements.

We also want to mention in passing that we have repeated the same 
calculation for  
single GRBs with fixed red shift ($z$) values with similar results.

\newpage
\begin{center}
{\bf Figure Caption}
\end{center}
\noindent {\bf Fig. 1} Variation of ${\cal R}$ with $\theta_{12}$ and 
$\theta_{23}$ for (a) $\theta_{13} = 0^o$, (b) $\theta_{13} = 3^o$,
(c) $\theta_{13} = 6^o$ and (d) $\theta_{13} = 9^o$. See text for details.

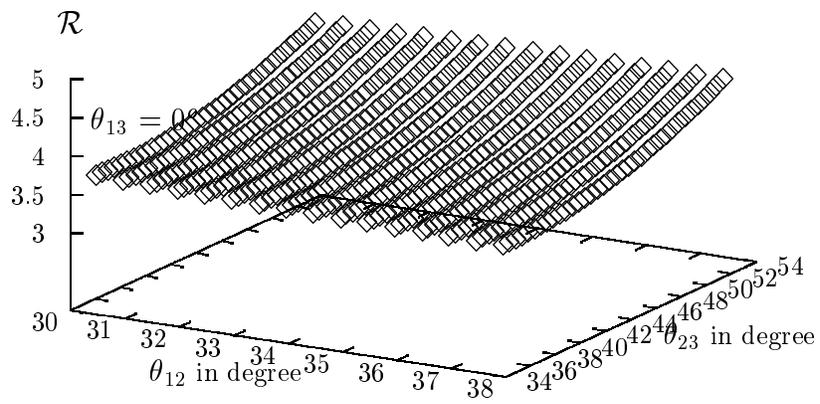
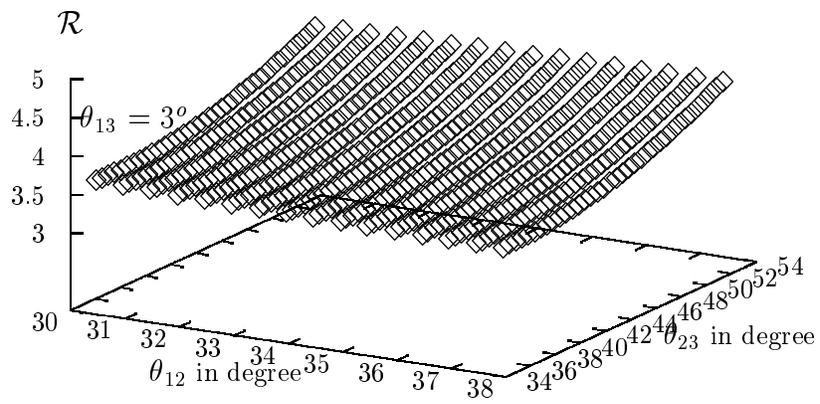
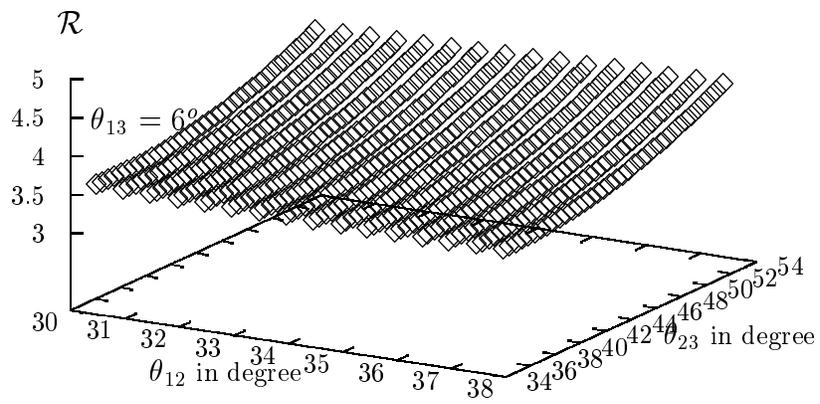
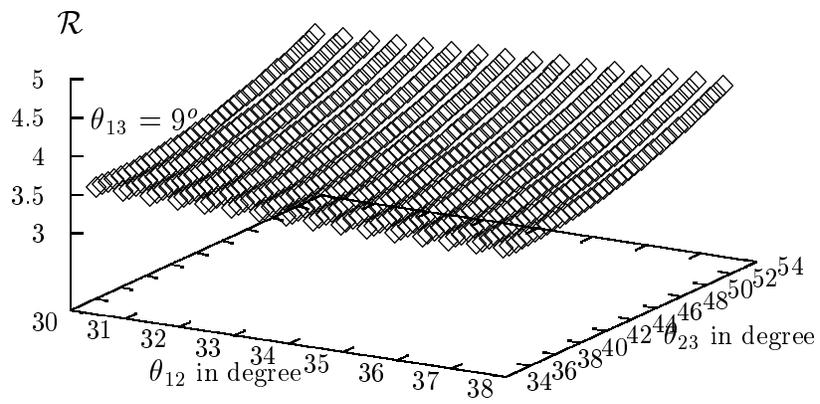
\end{document}